\documentclass[prl,reprint,superscriptaddress]{revtex4-2}

 \usepackage{amsmath}
 \usepackage{amsfonts,amssymb,marginnote}
 \usepackage{graphicx}
 \usepackage{bm}
 \usepackage{newtxtext,newtxmath}
\usepackage{dcolumn}

 \usepackage{hyperref}
 \hypersetup{
    colorlinks=true,       
    linkcolor=blue,          
    citecolor=blue,        
    filecolor=magenta,      
    urlcolor=blue           
}

\newcommand{\red}[1]{{\textcolor{red}{#1}}}

\renewcommand{\epsilon}{\varepsilon}

\allowdisplaybreaks

\newcolumntype{d}[1]{D{.}{.}{#1}}

\let\originalleft\left
\let\originalright\right
\renewcommand{\left}{\mathopen{}\mathclose\bgroup\originalleft}
\renewcommand{\right}{\aftergroup\egroup\originalright}

\usepackage{comment}

\usepackage[final]{pdfpages}
\makeatletter
\AtBeginDocument{\let\LS@rot\@undefined}
\makeatother

\begin{document}
\frenchspacing

\title{Many-Body Theory Calculations of Positron Scattering and Annihilation in H$_2$, N$_2$ and CH$_4$}
\author{C.~M. Rawlins$^{\dagger}$}
\affiliation{Centre for Light-Matter Interactions, School of Mathematics and Physics, Queen's University Belfast,\\ Belfast BT7 1NN, Northern Ireland, United Kingdom}
\author{J.~Hofierka$^{\dagger}$}
\affiliation{Centre for Light-Matter Interactions, School of Mathematics and Physics, Queen's University Belfast,\\ Belfast BT7 1NN, Northern Ireland, United Kingdom}
\author{B. Cunningham}
\affiliation{Centre for Light-Matter Interactions, School of Mathematics and Physics, Queen's University Belfast,\\ Belfast BT7 1NN, Northern Ireland, United Kingdom}
\author{C.~H. Patterson}
\affiliation{School of Physics, Trinity College Dublin, Dublin 2, Ireland}
\author{D.~G. Green}
\email{d.green@qub.ac.uk\\ ${\dagger}$ Joint-first authors.}
\affiliation{Centre for Light-Matter Interactions, School of Mathematics and Physics, Queen's University Belfast,\\ Belfast BT7 1NN, Northern Ireland, United Kingdom}
\date{\today}

\begin{abstract}
The recently developed \emph{ab initio} many-body theory of positron molecule binding [J. Hofierka \emph{et al.}, Nature, 606, 688 (2022)] is combined with 
the shifted pseudostates method [A. R. Swann and G. F. Gribakin, Phys.~Rev.~A {101}, 022702 (2020)] to calculate
positron scattering and annihilation rates on small molecules, namely H$_2$, N$_2$ and CH$_4$. 
The important effects of positron-molecule correlations are delineated. 
The method provides uniformly good results for annihilation rates on all the targets, from the simplest (H$_2$, for which only a sole previous calculation agrees with experiment), to larger targets, where high-quality calculations have not been available.
\end{abstract}

\maketitle

\setlength{\abovedisplayskip}{5.3pt}
\setlength{\belowdisplayskip}{5.3pt}

Developing fundamental knowledge of positron scattering and annihilation in molecules is essential to e.g., realize antimatter-based molecular spectroscopy \cite{RevModPhys.82.2557,antichem,Swann:2020b} and next-generation antimatter traps \cite{Danielson:2015,fajans:2020,Baker2021,Swann:2023}; elucidate the process of molecular fragmentation \cite{ionprod,posfrag,frag2,nrgdepo}; and
properly understand how positrons propagate in and can act as probes of living tissue (relating to DNA damage and dosimetry in PET \cite{Blanco2013,White:2014,Boyle:2015,Sanche:2000,DNA2011,PETbook}), the galaxy (e.g., to understand the galactic-centre annihilation signal \cite{e+fun,RevModPhys.83.1001} and dark matter \cite{Flambaum:2021}), and materials \cite{RevModPhys.66.841}.

The positron-molecule system is, however, characterized by strong positron-molecule correlations that are non-local and act over different length scales \cite{Hofierka2022}, and for molecules that bind the positron, spectacular resonance effects due to coupled electronic and vibrational dynamics \cite{RevModPhys.82.2557}. 
They make the theoretical and computational description a challenging many-body problem. 
For positron scattering, $R$-matrix \cite{Tennyson_1987,Baluja_2007,Zhang_2011,Edwards:2021,Graves2022}, Schwinger multichannel \cite{Germano_1993,SchwingerH2:wrong,Claudia_2000,d_A_Sanchez_2004,Zecca:2012,Oliveira_2012,Barbosa_17,Silva_20},  Kohn variational \cite{Armour_1989b, Armour_1990,COOPER_08, armourh2_1,Armour_2010}, model-potential \cite{Reid:2004,Swann_small}, and CCC (convergent close coupling) \cite{Zammit:2013,Zammit:2017} methods 
have been applied with considerable success to small molecules including H$_2$, CH$_4$, N$_2$, CO$_2$, CO, allene, formamide and pyrazine (see also \cite{Brunger:2017}).
Calculation of the positron-molecule annihilation rate --- of chief interest in this work --- is however, strikingly more difficult.
For a gas of number density $n_g$ the positron annihilation rate is parametrized as $\lambda=\pi r_0^2cn_gZ_{\rm eff}$, where $r_0$ is the classical electron radius, $c$ is the speed of light, and $Z_{\rm eff}$ is the effective number of electrons that participate in the annihilation process. 
Formally $Z_{\rm eff}$ is equal to the electron density at the positron, 
$\mathrm{Z_{eff}}  = \int \sum_{i=1}^{N_e} \delta(\bm{r}-\bm{r}_i) 
|\Psi_{\bf k}(\bm{r}_1, \dots, \bm{r}_{N_e}, \bm{r})|^2 d{\bf r}_1 \dots d{\bf r}_{N_e} d{\bf r}$,
where $\Psi_{\bf k}$ is the total wavefunction of the system, with electron coordinates ${\bf r}_i$ and positron coordinate ${\bf r}$ \footnote{For molecules the wavefunctions depends on the nuclear coordinates also. The molecules we consider here do not bind the positron, and we perform calculations in the fixed-nuclei approximation of the direct (non-resonant) annihilation rate.}. It describes the scattering of positron of momentum ${\bf k}$ by the molecule, and is normalised asymptotically to the product of the ground-state target molecular wavefunction and positron plane wave. Accurate calculation of $Z_{\rm eff}$ thus requires proper account of the scattering dynamics and positron-molecule correlations, including short-range electron-positron interactions. 
Even for the simplest molecule, H$_2$, calculations of $Z_{\rm eff}$ via sophisticated methods including R-matrix \cite{Zhang_2011} and the Kohn-variational \cite{armourh2_0,armourh2_1} and Schwinger multichannel methods \cite{VARELLA2002} disagree, all substantially underestimating experiment \cite{charlton_h2,larrichia_h2,Charlton_2013} (by $\sim$15--50\%), to which only a stochastic variational method calculation \cite{Zhang_Mitroy_2011}  is compatible. For N$_2$, used ubiquitously as a buffer-gas in positron traps \cite{Danielson:2015,fajans:2020}, the Schwinger multichannel method (the only \emph{ab initio} calculation we are aware of) underestimates experiment by a factor $>3$. Moreover, these methods cannot be easily scaled to larger molecules. Theoretical developments are demanded. 

Many-body theory is a powerful method that can accurately account for strong positron and electron  correlations with atoms, molecules and ions (see e.g., \cite{Amusia:Pos:MBT:He, dzuba_mbt_noblegas,ADC3,Cederbaum-elecpos,Cederbaum1996,Amusia_2003,Gribakin:2004,CCRMP,PhysRevLett.105.203401,DGG_posnobles, DGG:2017:ef,Amusia_2021,Safronova:MBT:elcAtom,DGG_hlike}). For atoms, a B-spline implementation provided a complete \emph{ab initio} description of positron scattering, annihilation and cooling \cite{DGG_posnobles,DGG:2015:core,DGG:2017:ef,DGG_cool,DGG_gamcool} and positronium (Ps) `pickoff' annihilation \cite{DGG:2018:PRL} in (noble-gas) atoms.
Most recently, we developed and successfully applied a multicentred Bethe-Salpeter Gaussian-orbital based many-body approach to positron binding in molecules, implemented in our {\tt EXCITON+} code \cite{Hofierka2022}.

Here, we extend the approach beyond binding, combining it with the recently-devised shifted-pseudostate-normalization method of Swann and Gribakin \cite{Swann_small} to perform fixed-nuclei \footnote{The molecules we consider do not bind a positron, and thus coupling of vibrational and electronic degrees of freedom is not required at this level of calculation.} calculations of low-energy positron scattering and (direct) annihilation rates on the
small molecules H$_2$ (the most abundant molecule in space), N$_2$ (a key buffer-gas in positron traps), and CH$_4$ (abundant in planetary atmospheres). We quantify the effects of positron-molecule correlations including positron-induced polarization, screening, virtual-Ps and positron-hole interactions, and compare with experiment and theory where available. For the annihilation rates, we find near perfect agreement with the benchmark stochastic variational calculation \cite{Zhang_Mitroy_09,Zhang_Mitroy_2011} and experiment for H$_2$, providing a consensus, 
and find excellent agreement with experiment for N$_2$ and CH$_4$. 

\begin{figure}[t!!]
\centering
\includegraphics[width=0.5\textwidth]{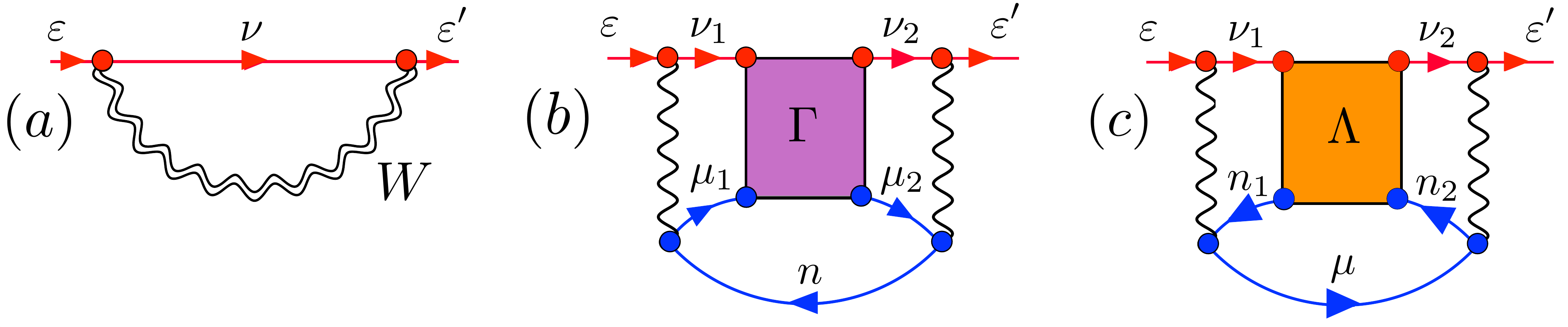}
\caption{\small The main contributions to the positron-molecule self energy:
{(a)}, the $GW$ diagram, which describes polarisation, and screening and electron-hole interaction corrections to it; 
(b) and (c),  the infinite ladder series of screened electron-positron interactions (`$\Gamma$-block') and positron-hole interactions (`$\Lambda$-block'). Lines labelled $\nu$ ($\mu$) (($n$)) are excited positron (electron) ((hole)) propagators; a single (double) wavy line denotes a bare (dressed) Coulomb interaction. See Extended Data Fig.~1 of \cite{Hofierka2022} for full details. \label{fig:diags}}
\end{figure}

\emph{Theory and numerical implementation.---}The positron (quasiparticle) wavefunction $\psi_{\epsilon}$ in the field of a many-electron target is found from the Dyson equation \cite{mbtexposed,Hofierka2022}:
\begin{eqnarray}\label{eqn:dyson}
\left(H^{(0)}+\hat{\Sigma}_{\epsilon}\right)\psi_{\epsilon}(\bm r) 
= \epsilon \psi_{\epsilon}(\bm r),
\end{eqnarray}
Here $H^{(0)}$ is the zeroth-order Hamiltonian, 
which we take to be that of the positron in the Hartree-Fock (HF) field of the ground-state molecule, and 
$\hat{\Sigma}_{\epsilon}$ is the non-local, energy-dependent correlation potential 
(self-energy 
\cite{BellSquires}).
In practice we calculate the matrix elements of $\Sigma$ via its diagrammatic expansion in the residual electron-electron and electron-positron interactions
\footnote{Since in Eqn.~(\ref{eqn:dyson}) $\Sigma_{\varepsilon}$ depends on the energy of the pseudostate involved, we first calculate $\Sigma_{E}$ on a dense energy grid and interpolate to the energy of the pseudostate (see Supplementary Information Fig.~1). }. 
See Ref.~\cite{Hofierka2022} for full details. 
Briefly, we include three classes of infinite series in the expansion: 
Fig.~\ref{fig:diags} (a) the `$GW$' diagram (the product of the positron Green's function $G$ and the screened Coulomb interaction $W$, which we calculate at the Bethe-Salpeter-Equation level), describes the positron-induced polarization of the molecular electron cloud, the screening of it by the molecular electrons, and electron-hole attractions; (b) the electron-positron ladder series (`$\Gamma$ block') that describes the non-perturbative virtual-Ps formation process and; (c), the positron-hole ladder series (`$\Lambda$ block'). 

We expand the electron and positron states in Gaussian basis sets (see below), transforming Eqn.~(\ref{eqn:dyson}) into a linear matrix equation. For a target that has no bound states for the positron, its solution yields a set of $n$ discrete positron continuum pseudostates and their corresponding energies $\varepsilon_n$ ($n=1,2,\dots$). These pseudostates decay exponentially rather than oscillate at large positron-target separations, and are normalized to unity instead of to an asymptotic plane wave as required by a true continuum state. Moreover, the lack of spherical symmetry of the multicentered target means that the orbital angular momentum is not conserved. However, at low positron momenta ($kR_a\ll 1$, where $R_a$ is the radius of the target), the mixing between partial waves due to the noncentral nature of the potential is small or negligible, and one can identify (approximately) states with 
eigenvalues of the squared orbital angular momentum operator $L^2$ close to zero ($s$-states), which are expected to dominate the low-energy scattering and annihilation.
In this case we can obtain the appropriate normalization following Swann and Gribakin \cite{Swann_small}, 
comparing the energies of (approximate) $s$-states 
against corresponding free positron pseudostate energies $\varepsilon_n^{(0)}$ (found by setting $H_0$ equal to the positron kinetic energy).
We thus calculate the $s$-wave phase shift for a positron of energy $\varepsilon_n$ as 
$\delta_0 = [n-f^{-1}(\varepsilon_n))]\pi$,
where $n$ is the number of the s-wave pseudostate, and $f(n)$ is a function of a continuous variable $n$ satisfying $f(n)=\varepsilon_n^{(0)}$ \cite{Swann_small}. We use the same procedure for $p$- and $d$-waves.
Moreover, we make use of the shifted energies to approximate the annihilation rate as $\mathrm{Z_{eff}}  = {4\pi}\delta_{ep}{A^{-2}}$,
with normalisation factor $A^2 = {2\sqrt{2\epsilon}}\,{\pi}{d\epsilon}/{dn}$ \cite{Swann_small} and 
$\delta_{\rm ep} = {2\sum_{i=1}^{N_e/2}}\gamma_i\int
|\varphi_i(\bm{r})|^2|\psi(\bm{r})|^2d\tau$
is the annihilation contact density summed over all occupied electronic MOs $\varphi_i$, including vertex enhancement factors $\gamma_i=1+\sqrt{{1.31}/{|\epsilon_i}|}+\left({0.834}/{|\epsilon_i|}\right)^{2.15}$ for MO $i$ with energy $\varepsilon_i < 0$ 
that account for the effects of short-range electron-positron Coulomb attraction \cite{DGG:2015:core,DGG:2017:ef}. 

We implement the above in our {\tt EXCITON+} Gaussian-basis code \cite{Hofierka2022} 
using aug-cc-pVXZ (X=T or Q) basis sets on the atoms of the molecule and up to 20 `ghost' centres away from the molecule to describe virtual-Ps formation, and a 19s17p16d15f even-tempered set on the molecular centre to help describe the long-range interactions; we assessed convergence and sensitivity to bond lengths (see Supplemental Material (SM) for full details)  \footnote{We verified the veracity of the code by reproducing B-spline based results for scattering on noble-gas atoms \cite{Hofierka:2023}.}.

\begin{figure}[t!]
\includegraphics*[width=0.23\textwidth]{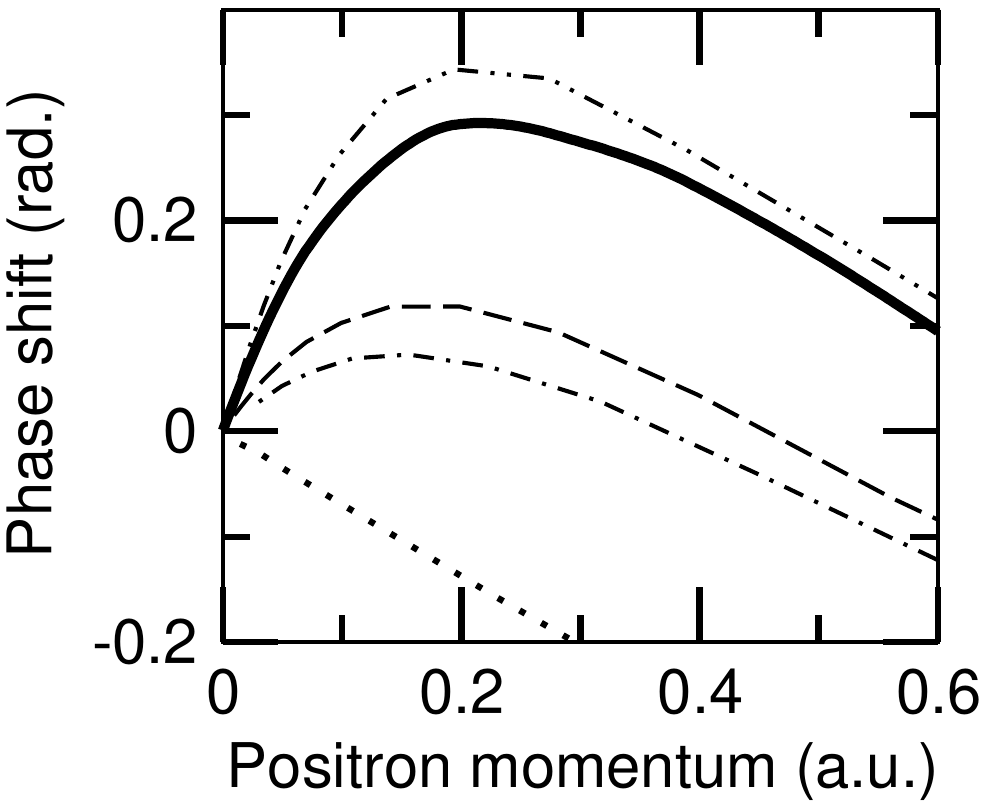}
\includegraphics*[width=0.23\textwidth]{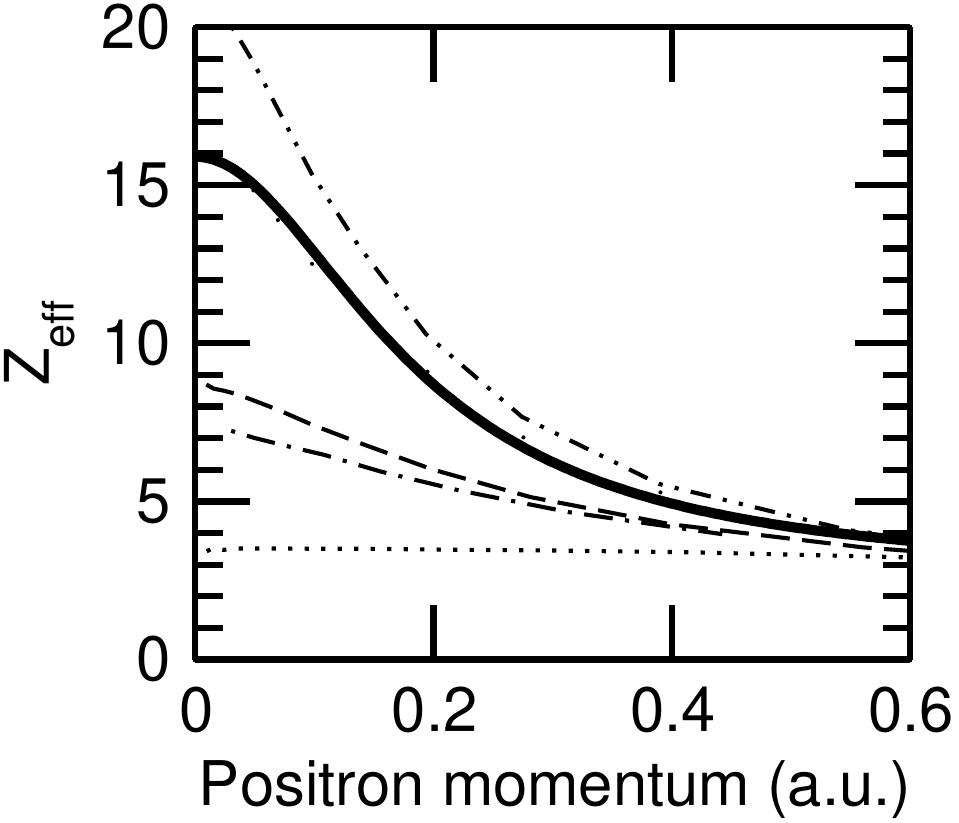}
\caption{
The effects of positron-molecule many-body correlations shown via the calculated $s$-wave scattering phase shift (left) and normalized annihilation rate $Z_{\rm eff}$ (right) for H$_2$ (representative of the three molecules considered in this work) in different approximations to the positron-molecule self energy (see Fig.~\ref{fig:diags}): Hartree Fock (black dotted); bare polarization 
$\Sigma^{(2)}$ (black dot-dashed); 
$GW$ (black dashed); 
$GW+{\Gamma}$ (black dot-dot-dashed); and 
$GW+{\Gamma}+{\Lambda}$ (solid line). 
\label{fig:phaseshifts}
}
\end{figure}

\begin{figure*}[ht!]
\includegraphics*[width=1.0\textwidth]{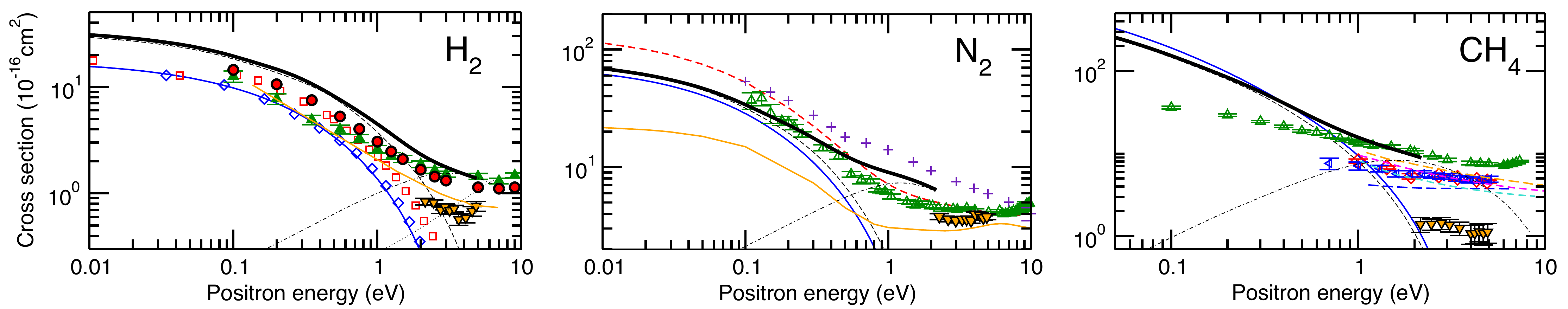}
\caption{\label{fig:res}
Many-body theory calculated scattering cross sections for H$_2$, N$_2$ and CH$_4$: $s$-wave (thin dashed black line), $p$ wave (thin dash-dotted black line), $d$-wave (dotted) and total (thick solid black line). 
Results are shown for bond lengths of $R=1.45$, 2.014, and 2.06 a.u. for H$_2$, N$_2$ and (the C-H bond in) CH$_4$.
Also shown are measurements by Zecca \emph{et al.} \cite{Zecca_2009, Zecca_2011, Zecca:2012} (green triangles) and Charlton \emph{et al.} \cite{Charlton_1983} (orange  triangles) for each molecule; the Schwinger multichannel \cite{PhysRevA.58.3502} (orange line), 
Kohn variational \cite{COOPER_08} (blue diamonds), convergent-close-coupling \cite{Zammit:2017} (red filled circles) calculations and
modified-effective-range-theory fit of measured cross sections \cite{Fedus_2015} (red squares)
for H$_2$; 
Schwinger multichannel  \cite{Claudia_2000} (orange line),
local complex potential \cite{EllisGibbings_2019} (plus symbols) 
and 
correlation-polarization-model \cite{Tenfen:2022} (dashed red) calculations  for N$_2$;
correlation-polarization-model-potential calculations of Franz \cite{Franz_2017} (orange dashed line), 
Jain and Gianturco \cite{Jain_1991} (blue dashed line),
Swann and Gribakin \cite{Swann_small} (blue solid line) and
Dibyendu \emph{et al.} \cite{Mahato_2021} (magenta dashed line), 
Schwinger multichannel  \cite{Zecca:2012} (turquoise dashed),
and measurements of
Sueoka and Mori \cite{Sueoka_1986} (blue triangle) and 
Dababneh \emph{et al.} \cite{Dab:1988} (red diamonds) for CH$_4$. 
\label{fig:xsecs}}
\end{figure*}

\begin{figure*}[t!]
\includegraphics*[width=1.0\textwidth]{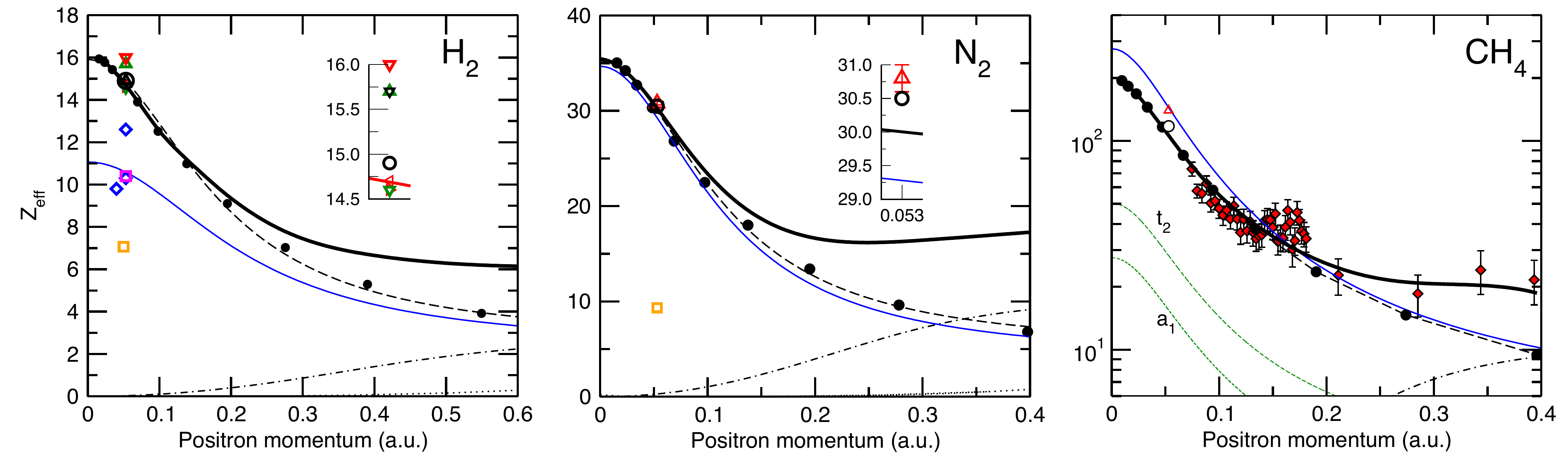}
\caption{Many-body theory calculated annihilation rate $Z_{\rm eff}(k)$ (using the $GW+\Gamma+\Lambda$ self-energy and including annihilation vertex enhancement factors) for H$_2$, N$_2$ and CH$_4$: $s$-wave (thin dashed black line), $p$-wave (thin dashed-dotted), total $s+p+d$ (thick solid black line). Results are shown for bond lengths of $R=1.4$, 2.014, and 2.06 a.u. for H$_2$, N$_2$ and (the C-H bond in) CH$_4$.
Also shown are the room-temperature Maxwellian averaged $\bar{Z}_{\rm eff}$ from our calculation (black open circle; for H$_2$ black down triangle is for $R=1.45$ a.u. for comparison) and experiment (red triangles) for H$_2$ \cite{PhysRevA.20.347, charlton_h2,larrichia_h2}, N$_2$ \cite{Charlton_2013}, and CH$_4$ \cite{Charlton_2013}, along with energy-resolved measurements for CH$_4$ \cite{MARLER200484} (red diamonds).
For H$_2$ we also show the calculated room temperature values from the Kohn variational method (blue diamonds) \cite{armourh2_0, armourh2_1,  armourh2_new}, molecular R-matrix (magenta square) \cite{Zhang_2011} and stochastic variational method at bond length of $R=1.4$ a.u. (green triangle down) and  $R=1.45$ a.u. (green triangle up) \cite{Zhang_Mitroy_09}, and the Schwinger multichannel method at $k=0.05$ a.u. (orange squares) \cite{Claudia_2000,VARELLA2002}. The latter is also shown for N$_2$. 
For CH$_4$ we also show the individual s-wave contributions from the $2a_1$ and one of the triply-degenerate $t_2$ HOMOs). 
\label{fig:zeff}}
\end{figure*}

\emph{Results: effect of many-body correlations.---}Figure \ref{fig:phaseshifts} shows the phase shifts and normalized annihilation rate $Z_{\rm eff}$ for H$_2$ (representative of the three molecules considered) calculated at the Hartree-Fock (HF), $\Sigma^{(2)}$, $GW$, $GW+\Gamma$, and $GW+\Gamma+\Lambda$ level for the correlation potential (see Fig.~\ref{fig:diags}). 
At the HF level the positron-molecule interaction is repulsive (corresponding to a negative phase shift, and small electron-positron overlap and thus annihilation rate); including the bare polarization $\Sigma^{(2)}$ \footnote{$\Sigma^{(2)}$ is found from Fig.~1 (a) with $W$ approximated as $V\Pi^{(0)}V$, where $\Pi^{(0)}$ is the HF electron-positron two-particle propagator). See \cite{Hofierka2022} Extended Fig.~1.} produces an attractive interaction at low momenta (turning the phase shift positive and increasing the electron-positron overlap and thus $Z_{\rm eff}$), which is further enhanced by the inclusion of the dressed ring diagrams of the $GW$@BSE, i.e., the intra-ring BSE electron-hole attractions are larger than the repulsive screening effects from the random-phase approximation ring series. The additional inclusion of the virtual-Ps contribution ($GW+\Gamma$) further increases the attractive potential substantially, causing a factor of $\sim 3$ increase in the phaseshift maximum and a more than doubling of $Z_{\rm eff}$ at low momenta, but is tempered by the repulsive positron-hole ($\Lambda$-block) contribution. 
The corresponding graphs for N$_2$ and CH$_4$ are presented in Fig. S4 in the Supplemental Material. We now consider the scattering cross sections and annihilation rates, comparing with experiment and theory where available.

\emph{Cross sections.---}
Figure \ref{fig:xsecs} shows the calculated elastic cross sections.
For H$_2$, for which the various calculations agree to $\lesssim10$\% error, our calculation agrees best with (though is slightly larger than) the accurate convergent-close-coupling calculation \cite{Zammit:2017}, the model-potential calculation \cite{Swann_small} (which underestimates the $Z_{\rm eff}$, see below)
 and measurements of Zecca \emph{et al.} \cite{Zecca_2009}. 
 For N$_2$ our result is compatible with the measurements of Zecca \emph{et al.} \cite{Zecca:2012}. For CH$_4$ our results  are consistent with the measurements of Zecca \emph{et al.} \cite{Zecca:2012} at $\sim 1$ eV, but are considerably larger at low energies (where they are compatible with the model-potential calculation \cite{Swann_small} that gives $Z_{\rm eff}$ in agreement with experiment, see below). 
We also calculate the scattering length $a$ from fits of the effective range theory expansion to the calculated phase shifts (see Supplemental Material Table I), finding them to be within $\sim$5-20\% error of previous calculations. 

\emph{Annihilation rate $Z_{\rm eff}$.---}Of chief interest in this work is the annihilation rate $Z_{\rm eff}$, due to the challenge it poses for theory and lack of accurate methods. Figure \ref{fig:zeff} shows our normalized annihilation rate $Z_{\rm eff}(k)$ as a function of positron momentum calculated in our most sophisticated approximation ($GW+\Gamma+\Lambda$ self-energy and including vertex enhancement factors). 
We show the discrete data points calculated for the $s$-wave, along with fits to the physically motivated form \cite{DGG_posnobles}
$Z_{\rm eff}(k) = F/({\kappa^2+k^2+Ak^4})+B$ where $F, \kappa, A$ and $B$ are constants \footnote{($F, \kappa, B$ and $A$) were found to be
(0.45, -0.184, 2.612, 0.002) and 
(0.454, -0.180, 2.719, -0.003)
for H$_2$ using bond length $R=1.4$ and $R=1.45$ a.u. respectively;
(0.407, -0.116, 4.982, -0.007) and
(0.411, -0.112, 5.075, -0.001)
for N$_2$ using bond length $R=2.014$ and $R=2.068$ a.u. respectively;
and (0.57, -0.056, 8.26, -0.03) for CH$_4$.
}.
We also show the calculated room-temperature Maxwellian average (open circle)
$\bar{Z}_{\rm eff} 
=
(2\pi k_BT)^{-3/2}\int^{\infty}_0 Z_{\rm eff}(k)
{{\rm exp}(-k^2/2k_BT)}{}\,
4\pi k^2dk.$
Table \ref{Zeff_fit_table} gives the values of $\bar{Z}_{\rm eff}$: for H$_2$ and N$_2$ (CH$_4$), we found it to be $<1\%$ (10\%) larger than  $Z_{\rm eff}(k)$ at thermal $k\sim0.05$ a.u.

\begin{table}
\caption{Maxwellian-averaged annihilation rate $\bar{Z}_{\rm eff}$.} 
\begin{tabular}{l@{\hspace{4pt}}lll}
\hline\\[-2ex]
\hline\\[-2ex]
& H$_2$ & N$_2$ & CH$_4$\\[.1ex]
\hline\\[-1ex]
{\bf Present MBT}\footnote{Positron-molecule self energy at $GW+\Gamma+\Lambda$ [Fig.~1 (a)+(b)+(c)].}
& 14.9, 
15.7\footnote{H$_2$ calculation using bond lengths of $R = 1.40$ a.u., $R = 1.45$ a.u.} 
& 30.5\footnote{N$_2$ calculation using bond lengths of  $R = 2.014$ a.u.}
& 119.2\footnote{CH$_4$ calculation using C-H bond length of  $R = 2.06$ a.u.,} \\[0.6ex]

SMC\footnote{Schwinger multichannel method at $k=0.05$ a.u.} \cite{Claudia_2000,VARELLA2002} & 7.70 & 8.96 & --\\
R-matrix  \cite{Zhang_2011} &10.4 &-- & --\\ 
Kohn var. \cite{armourh2_0,armourh2_1} & 12.6\footnote{Kohn-variational `method of models' calculation.} & -- &--\\
SVM \cite{Mitroy:H2} & 14.6, 15.7\footnotemark[2] & -- &--\\
Corr. pol. \cite{GIANTURCO200017,Surko:2005}& -- & $44\pm4$ & 99.5 \footnote{Correlation polarization potential calculations.}\\

LCAO\footnote{Linear-combination-of-atomic-orbital with correlation adjustment factors.} \cite{DGG_molgamma} & 14.6 & -- & -- \\
Model-pot. 
\cite{Swann_small} & 10.6 & 29.8 & 163 \\[0.5ex]
{\bf Experiment} 
&  14.7 $\pm$ 0.2 \cite{PhysRevA.20.347}
& 30.8$\pm$0.2\cite{Charlton_2013} & $140\pm0.8$\cite{Charlton_2013}\\
&  {16.0} $\pm$ 0.2 \cite{charlton_h2,Charlton_2013}\footnote{Ref.~\cite{Charlton_2013} recommended value.}  \\
& 14.6 $\pm$ 0.1 \cite{larrichia_h2}\\
\hline
\end{tabular}
\label{Zeff_fit_table}
\end{table}

Considering comparison with other theory and experiment, for H$_2$ a number of sophisticated calculations of $Z_{\rm eff}$, namely the Schwinger variational (7.7), R-matrix (10.4) and Kohn-variational (12.6), are in considerable disagreement, and moreover, all substantially underestimated experiment (14.6 -- 16 \cite{charlton_h2,Charlton_2013,PhysRevA.20.347,larrichia_h2}, with 16 the recommended value \cite{Charlton_2013}). The only compatible calculation to date is the stochastic variational calculation of Zhang and Mitroy \cite{Zhang_Mitroy_2011} (14.6 for a bond length of $R=1.4$ a.u., and 15.7 for a bond length of $R=1.45$ a.u. Our respective calculations of 14.9 and 15.7 are in near perfect agreement with the stochastic variational method and the experiment, providing a consensus, and demonstrate that the many-body theory accurately describes the correlations. The scattering length $a\sim1/2\kappa=-2.72$ determined effectively from the fit to $Z_{\rm eff}(k)$ is in $<$10\% error compared to the accurate convergent close coupling and stochastic variational calculations (see Supplemental Material Table I).

For N$_2$, the only \emph{ab initio} calculation we are aware of is the Schwinger multichannel calculation, which finds $Z_{\rm eff}=8.96$ \cite{Claudia_2000,VARELLA2002}, compared to the recommended measured value of 30.8$\pm0.2$ \cite{Charlton_2013}. In contrast, our calculated value of 30.5 is in excellent agreement with experiment, indicating proper account of the correlations that act to enhance $Z_{\rm eff}$, and with the recent model-potential calculation \cite{Swann_small}. We found that a 2\% increase in the bond length leads to a $\sim 5\%$ increase of $\bar{Z}_{\rm eff}$. We found that the fractional contribution to the $s$-wave $Z_{\rm eff}$ from the highest 5 MOs $a_{1g}a_{2u}a_{1g} 2e_u$ of HF ionization energies 40.67\,eV, 20.91\,eV, 17.34\,eV, 17.10\,eV and 17.10 eV to be 0.06, 0.24, 0.28, 0.21, and 0.21, respectively, i.e., a non-negligible fraction of annihilation occurs on MOs below the HOMO due to their favourable overlap with the positron wavefunction that is maximum around the N atoms. This pattern was observed for bound states \cite{Hofierka2022} and in the extensive fragmentation patterns \cite{ionprod,posfrag,frag2,nrgdepo} of polyatomic molecules.

For CH$_4$, we find excellent agreement with the positron-momentum-dependent $Z_{\rm eff}(k)$ measurements of Marler \emph{et al.} \cite{MARLER200484} across the full momentum range including up to $k\sim 0.4$ a.u., where the $p$-wave contributes (though we do not resolve the structure around 0.17 a.u.). Our thermalized value $\bar{Z}_{\rm eff}=119.2$ is lower than the measurement 140$\pm$0.8 \cite{charlton_h2,Charlton_2013}, and the model-potential calculation \cite{Swann_small} (which uses adjustable parameters), especially at small $k$. 
We found the fractional contribution from the 1$a_1$, 2a$_1$ and each of the $t_2$ orbitals, of ionzation energies 304.92\,eV, 25.66\,eV, 14.83\,eV, to be 0.0025, 0.15, 0.281. The large scattering length in CH$_4$ makes $\bar{Z}_{\rm eff}$ very sensitive to the correlation potential strength at low momenta (since $\kappa\sim 1/2a\ll1$). We assessed convergence of the basis set, increasing from 12 to 20 ghosts, and from TZ to QZ functions, finding only a 5\% increase (see SM Figure S2). 
We include angular momenta functions up to $\mathcal{\ell}=4$: whilst the basis functions from different centres combine to provide effectively higher angular momenta \cite{Swann:2018}, 
this may be insufficient to converge the virtual-Ps diagram \footnote{We only have access to modest computational resources, and the code is in its infancy. Larger resources, and optimisations of the code (including exploiting the point group symmetry) would enable larger calculations}.
Moreover, the annihilation-vertex enhancement factors (determined from \emph{ab initio} calculations for atoms \cite{DGG:2015:core,DGG:2017:ef}) may underestimate the true short-range enhancement for delocalized MOs, especially since the positron can probe electron density in interstitial regions where the nuclear repulsion is reduced \footnote{We calculate (at the $GW$ level) the vertical ionization energy of the HOMO as 14.8 eV, which is close to the 14.35 eV of experiment and ionization-energy-optimised $GW$ calculations (14.14-14.5 eV) \cite{vanSetten2015,Caruso2016}: using the experimental value would increase the enhancement factor by only 1.04, insufficient to bring theory into agreement with experiment.}. Calculation of the vertex enhancement for molecules is extremely challenging, and beyond the scope of this work.
Further theoretical and experimental work on CH$_4$ is warranted.

\emph{Summary and Outlook.---}
The accurate \emph{ab initio} calculation of the positron-molecule annihilation rate has proved to be a formidable problem, thwarting the efforts of quantum chemistry methods for all but the simplest molecule, H$_2$, for which only a sole (stochastic variational) calculation agrees with experiment. 
In this work, many-body theory was developed and applied to calculate positron scattering properties and annihilation rates in H$_2$, N$_2$ and CH$_4$.
The effects of correlations were elucidated. For the annihilation rates $Z_{\rm eff}$, for H$_2$, the power of the approach was demonstrated by reproducing the benchmark stochastic variational result, thus providing a consensus with experiment. Moreover, overall excellent agreement with experiment was also found for N$_2$ and CH$_4$ (though further theoretical and experimental work on the latter was called for).

The positron-molecule correlation potential the many-body approach provides can be by incorporated in a 
$T$-matrix (see e.g., \cite{Schneider:1970,McCurdy_1976,Klonover_1979,Berman:1983,Cederbaum1996} or Schwinger multichannel \cite{Germano_1993,SchwingerH2:wrong,Claudia_2000,d_A_Sanchez_2004,Zecca:2012,Oliveira_2012,Barbosa_17,Silva_20} approach to enable calculations on larger molecules, and should provide uniformly good accuracy.
The framework provides a foundation for a many-body description of annihilation $\gamma$ spectra, and of inelastic scattering (including Ps formation) \cite{Csanak:1971,Cederbaum1996,Cederbaum:2000}, 
and to describe the coupled electronic and vibrational degrees of freedom \cite{Brand:1999} (we note in this regard the impressive recent success of the close-convergent-coupling approach in electron scattering on H$_2$ \cite{Scarlett:2021}) for \emph{ab initio} calculation of resonant annihilation and scattering in positron-binding molecules \cite{RevModPhys.82.2557}. 
 
\begin{acknowledgments}
\emph{Acknowledgements.---}We thank Mike Charlton, Cliff Surko, James Danielson, Jack Cassidy and Sarah Gregg for useful discussions, and Gleb Gribakin and Andrew Swann for useful comments on the manuscript.  
This work was supported by the European Research Council grant 804383 ``ANTI-ATOM'', 
and used the Northern Ireland High Performance Computing service funded by EPSRC (EP/T022175)
and the ARCHER2 UK National Supercomputing Service. 
\end{acknowledgments}


%

\newpage
\newpage


\section*{Supplementary material}

# Supplemental Material
# Many-body theory calculations of positron scattering and annihilation in $H_2$, $N_2$ and $CH_4$

C. M. Rawlins[†],[1] J. Hofierka[†],[1] B. J. Cunningham,[1] C. H. Patterson,[2] and D. G. Green[1]

[1]*School of Mathematics and Physics, Queen's University Belfast, BT7 1NN, Northern Ireland, United Kingdom*
[2]*School of Physics, Trinity College Dublin, Dublin 2, Ireland*
(Dated: March 3, 2023)

## NUMERICAL IMPLEMENTATION: DETAILS OF BASIS SETS

We construct the self energy and solve the Dyson equation in our `EXCITON+` code [1] [2]. We use aug-cc-pVQZ Dunning basis sets [3, 4] centred on the atoms, enabling accurate determination of the electronic structure including polarizabilities. To more accurately describe the long-range positron-induced polarisation we place an additional basis set for the electron and positron at the molecular centre: for $H_2$ and $N_2$ we use an aug-cc-pVTZ basis set of hydrogen for the electronic basis and a diffuse even-tempered 19s17p16d15f basis for the positron, with exponents for the $j$-th Gaussian for each angular momentum $l$ given as $\zeta_j = \zeta_0 \beta^{j-1}$, with $\beta = 2$, $\zeta_0 = 10^{-5}$ for $l = 0 - 1$ and $\zeta_0 = 10^{-4}$ for $l = 2 - 3$. For $CH_4$ we place the diffuse even-tempered basis on the central carbon atom. Moreover, to more accurately describe the virtual-Ps formation process, which takes place a few a.u. away from the molecule and requires large angular momentum (to resolve the electron-positron distance and describe excited Ps), we place additional aug-cc-pVTZ basis sets of hydrogen for both electron and positron functions on 'ghost' centres surrounding the molecule. For $H_2$ and $N_2$ we used 12 on the vertices of a regular icosahedron with a circumscribed sphere of radius 2 and 3 a.u. from the centre of the molecule. For $CH_4$ we additionally considered 20 ghosts placed on the vertices of a regular dodecahedron at 3 a.u. distance from the centre of the molecule (shown in Fig. S1). Moreover, for $CH_4$ we also considered an alternative square-pyramidal cap arrangement, with each ghost 1Å away from the hydrogens, giving only a 0.5% improvement. Finally, we also upgraded the aug-cc-pVTZ basis set on the 20 ghost centres to QZ with only about 1% increase in the $\tilde{Z}_{eff}$ result. Convergence w.r.t. number of virtual states was assessed: see e.g., Supplemental Figure S2, which shows the convergence of $Z_{eff}$ and $\tilde{Z}_{eff}$ as a function of the electron and positron basis sizes for $N_2$ and $CH_4$.

## EFFECTS OF CORRELATIONS ON PHASE SHIFTS, ELASTIC SCATTERING CROSS SECTIONS AND $Z_{eff}$.

The effects of the various diagrams describing the positron-molecule correlations, similar to Fig .1 of the main text, is shown for the phase shifts, cross sections and $Z_{eff}$ for all of the molecules. One point to note is that for $H_2$ and $CH_4$ the effect of the attractive electron-hole interactions within the bubbles, which are included in the $GW$@BSE calculation dominates over the repulsive electron-screening for $N_2$, meaning that the correlation potential is more attractive compared to the bare self energy $\Sigma^2$. The opposite is found in $N_2$.

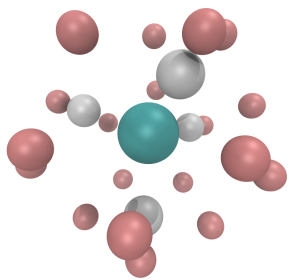

Supplemental Fig. 1. $CH_4$ showing the positions of the atoms and 20 'ghost' centres (coloured light red) on which we place basis sets. The ghost centres are placed on the vertices of a regular dodecahedron with a circumscribed sphere of radius 3 a.u.

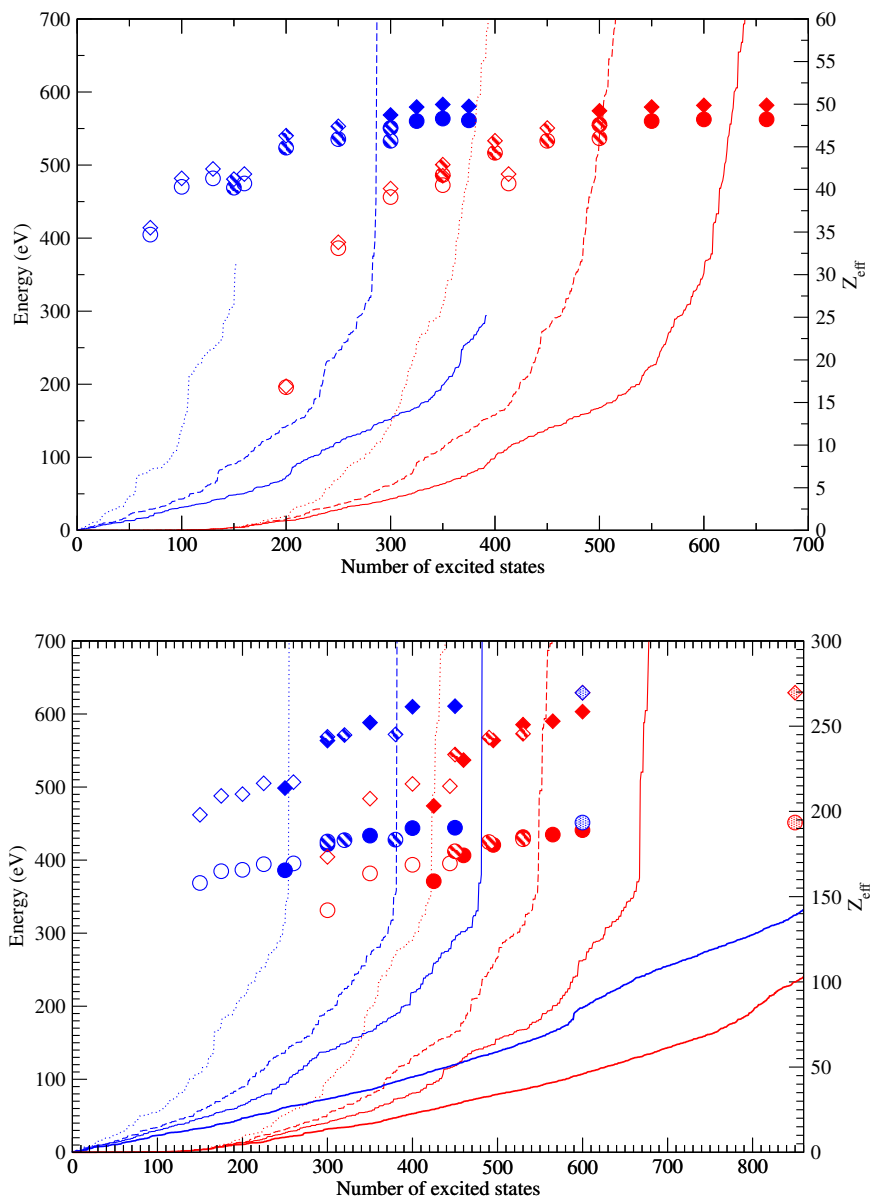

Supplemental Fig. 2. Convergence of $Z_{eff}$ in $N_2$ (top) and $CH_4$ (bottom) with respect to electron and positron basis sizes. Values of $Z_{eff}$ ($k$ = 0.053) (circles) and $\bar{Z}_{eff}$ (diamonds) calculated with the positron-molecule self energy at the $GW@BSE+\Gamma$ level, i.e., including only Fig. 1 (a) and Fig. 1(b) from the main paper; convergence of the $\Lambda$ block is faster) are shown with respect to the number of electron (blue) and positron (red) Hartree-Fock molecular orbitals included in the basis. The number of electron (positron) states is kept at maximum when varying positron (electron) basis size. Empty symbols and dotted lines are results of calculations using a single ghost atom (at the centre of mass), striped symbols and dashed lines are from calculations with 6 additional ghost atoms (3 a.u. away from center of mass) and solid lines and symbols are from calculations with 12 ghost atoms (with hydrogen aug-cc-pVTZ basis). For $CH_4$ an additional result using 20 ghost basis centres with hydrogen aug-cc-pVQZ basis functions is shown as dotted symbols ($Z_{eff}$) and thick lines (energies of the states). The final calculation used 850 positron states and 595 electron states. This corresponds to diagonalising a symmetric matrix of dimension 505,750, and was performed using 40 high-memory nodes of ARCHER2 using (a maximum of) 17 TB RAM.

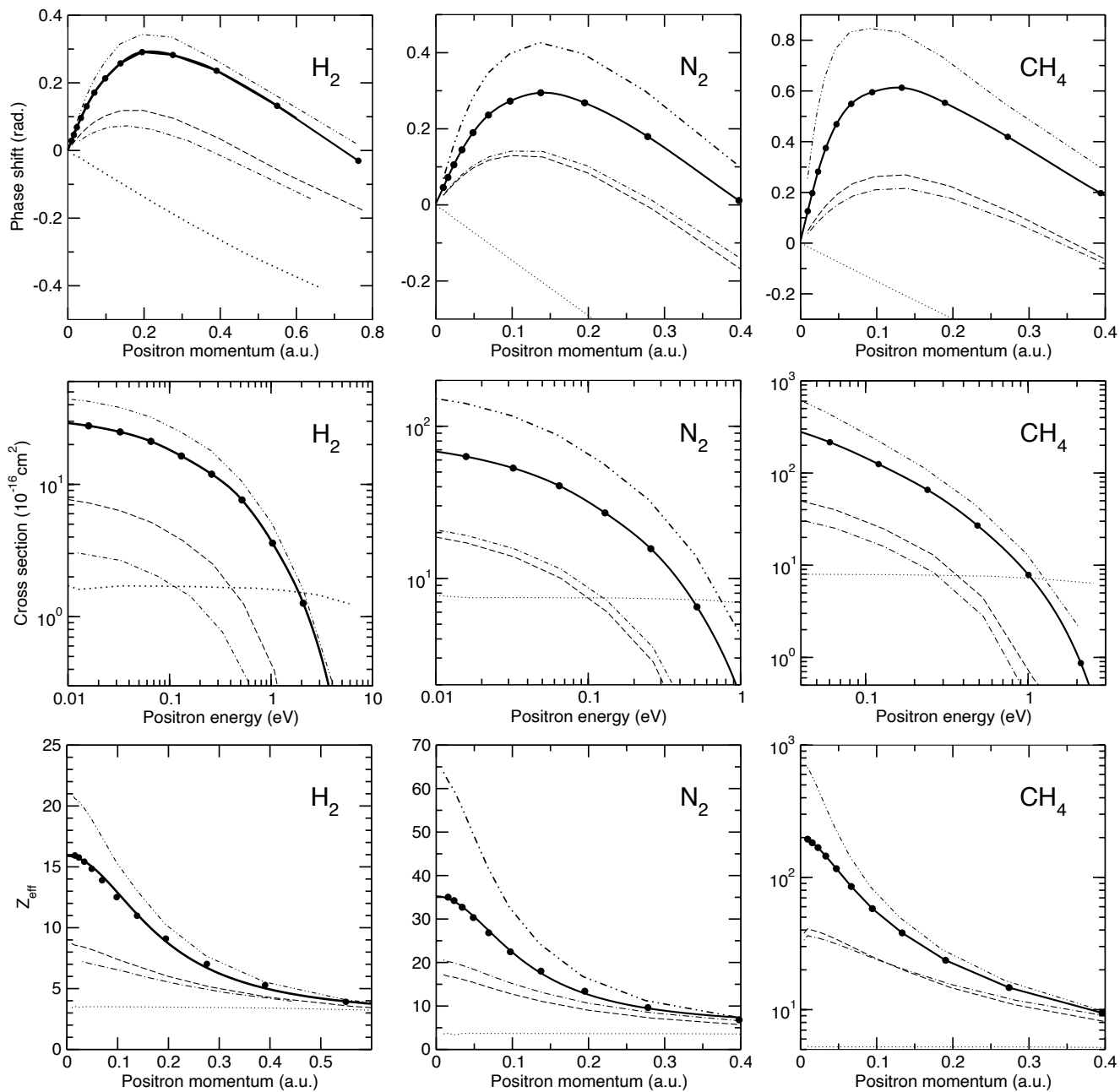

Supplemental Fig. 3. Effect of positron-molecule correlations. Calculated $s$-wave phase shifts, cross sections and normalized annihilation rate $Z_{eff}$ for $H_2$, $N_2$, and $CH_4$ in different approximations to the positron-molecule self energy: Hartree Fock (black dotted line); many-body theory at the $\Sigma^{(2)}$ (black dot-dashed); $GW$ (at BSE level), black dashed; $GW + \Gamma$ (black dot-dot-dashed); and $GW + \Gamma + \Lambda$ levels (thick solid line; black filled circles denote the discrete data. For the $Z_{eff}$ the solid lines are the fit to the the functional form described in the main text.

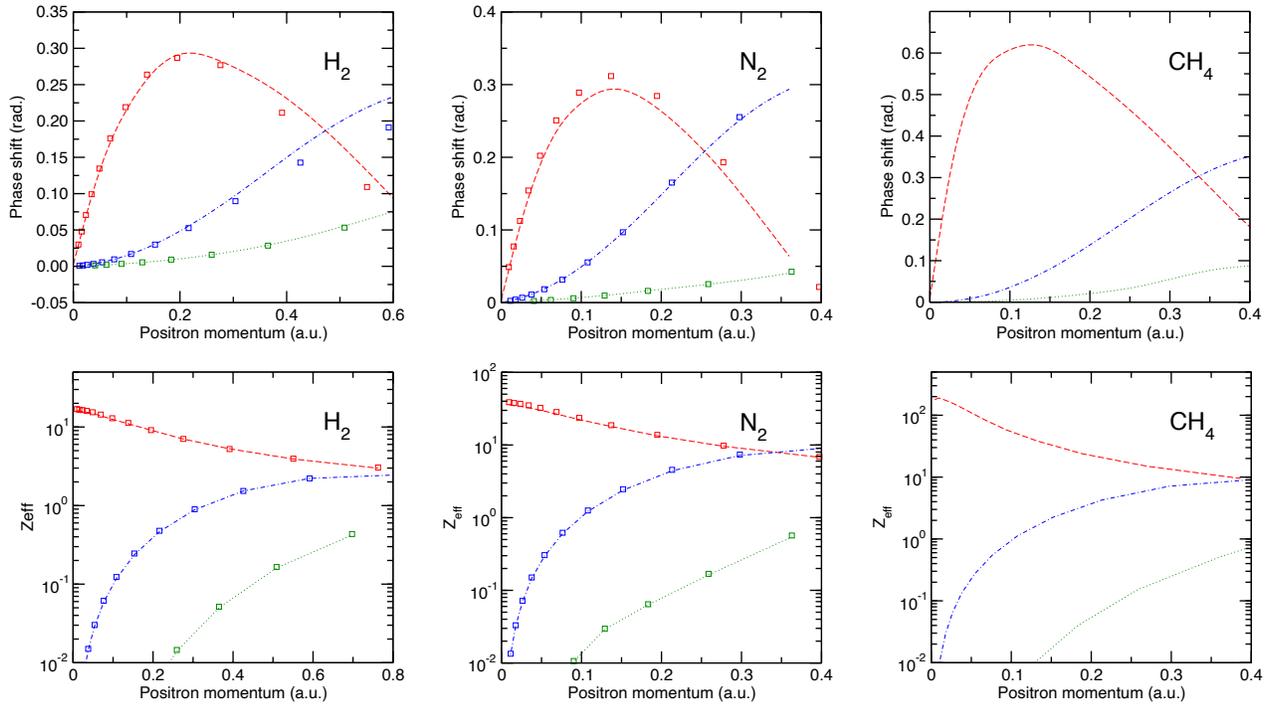

Supplemental Fig. 4. Partial wave contributions and sensitivity to the bond lengths. Calculated $s$ (red), $p$ (blue) and $d$-wave (green) phase shifts and normalized annihilation rate $Z_{\text{eff}}$ for H$_2$, N$_2$, and CH$_4$. For H$_2$ calculations were performed using a bond length of $R = 1.40$ (lines) and $R = 1.45$ a.u. (symbols). For N$_2$ calculations were performed for bond lengths of $R = 2.014$ a.u. (lines) and $R = 2.068$ a.u. (symbols).

## SCATTERING LENGTHS

In Supplemental Table I we present the calculated scattering length $a$ determined by fitting the $k \to 0$ phase shifts to the effective-range-theory expansion [5]

$$\tan \delta_0 = -ak \left[ 1 - \frac{\pi \alpha k}{3a} - \frac{4 \alpha k^2}{3} \ln \left( C \frac{\sqrt{\alpha} k}{4} \right) \right]^{-1} \qquad (1)$$

where $\alpha$ is the molecular dipole polarizability (calculated here as 5.05, 11.08 and 15.72 for H$_2$, N$_2$ and CH$_4$, respectively).

Supplemental Table I. Calculated scattering lengths (a.u.) from Eqn. (1).

|  | Eqn. (1) | $a \sim 1/2\kappa$[a] | Other calculations |
|---|---|---|---|
| H$_2$ | -3.07 | -2.72 | -2.39[b], -2.49[c], -2.57[d], -2.06[e], -2.2[f] |
| N$_2$ | -4.89 | -4.31 | -4.6±0.1[b], -4.64[g] |
| CH$_4$ | -13.4 | -8.9 | -17±1[b], -7.4[h], -13.0[i] |

[a] Determined effectively from fit to the annihilation rate $Z_{\text{eff}}$ (see text).
[b] Model potential [6].
[c] CCC (Convergent Close Coupling) calculation [7].
[d] Stochastic variational calculation [8].
[e] Molecular R-matrix with pseudostates [9].
[f] Kohn-variational 'method of models' calculation [10, 11].
[g] Modified effective range [12].
[h] Schwinger multichannel [13].
[i] Semi-empirical [14].



\end{document}